\newif\ifconf
\conftrue

\newif\ifcomm
\commtrue

\newif\ifblind
\blindfalse

\newif\ifs
\sfalse

\newif\ifhyperref
\hyperreftrue

\newif\ifacm
\acmfalse

\newif\ifusenix
\ifacm
   \usenixfalse %
\else
    \usenixfalse %
\fi

\newif\ifhotnets
\ifacm
   \hotnetsfalse %
\else
    \ifusenix
        \hotnetsfalse %
    \else
        \hotnetsfalse %
    \fi
\fi
    
\newif\ifacmart
\ifacm   %
   \acmarttrue
\else
   \acmartfalse
\fi

\newif\ifieee
\ifacm
   \ieeefalse
\else
    \ifusenix
        \ieeefalse
    \else
        \ifhotnets
            \ieeefalse
        \else
            \ieeetrue
        \fi
    \fi
\fi

\ifconf
    \newcommand{\Conf}[1]{#1}
    \newcommand{\TR}[1]{}
    \newcommand{\Journal}[1]{}  %
    \newcommand{\OnlyTR}[1]{}   %
\else
    \newcommand{\Conf}[1]{}
    \newcommand{\TR}[1]{#1}
    \newcommand{\Journal}[1]{#1}  %
    \newcommand{\OnlyTR}[1]{}   %
\fi

\ifacm
	\ifacmart
	    \ifblind
            \documentclass[sigconf,anonymous,nonacm]{acmart}
        \else
            \documentclass[sigconf]{acmart}
        \fi
    \else
      \Conf{
 	     \documentclass[letterpaper,conference]{sig-alternate-10pt}
      }
      \TR{
     	 \documentclass[letterpaper,conference]{sig-alternate}
      }
      \usepackage[
       pass,%
      ]{geometry}
	\fi
\else
	\ifusenix  %
		\documentclass[letterpaper,twocolumn,10pt]{article}
        \usepackage{usenix-2020-09}
        \usepackage{filecontents}
    \else %
	    \ifhotnets  %
            \documentclass{hotnets23}
    	\else  %
            \Conf{
                  \documentclass[conference]{IEEEtran}
            }
            \TR{
                  \documentclass[journal]{IEEEtran}
            }
        \fi
    \fi
\fi

\ifieee %
    \usepackage[noadjust]{cite} %
	\usepackage[cmex10]{amsmath}
	\usepackage{amsthm}  %
    \newtheoremstyle{boldthm}{}{}{\itshape}{}{\bfseries}{.}{ }{\thmname{#1}\thmnumber{ #2}\thmnote{ (#3)}} %
	\theoremstyle{boldthm}
\fi

\ifacmart 
\else 
    \ifhotnets
        \usepackage{times}  
        \ifhyperref
            \usepackage[hidelinks]{hyperref}
            \hypersetup{pdfstartview=FitH,pdfpagelayout=SinglePage}
        \fi

    \else
        \ifhyperref
            \usepackage[hidelinks]{hyperref} %
            \hypersetup{pdfstartview=FitH,pdfpagelayout=SinglePage}
        \fi
    \fi
\fi

\ifacmart
\else
  \newtheorem{theorem}{Theorem}%

\fi

\ifieee %
    \newcommand{\bp}{\begin{IEEEproof}}     %
    \newcommand{\bpo}{ \begin{IEEEproof}[Proof Outline] }
    \newcommand{\ep}{\end{IEEEproof}}       %
    \newcommand{\proofof}[1]{\begin{IEEEproof}[Proof of #1]} %
\else
    \newcommand{\bp}{\begin{proof}}
    \newcommand{\bpo}{ \begin{proof}[Proof Outline] }
    \newcommand{\ep}{\end{proof}}       %
    \newcommand{\proofof}[1]{\begin{proof}[Proof of #1]} %
\fi

\ifieee
    \usepackage[justification=justified, font=footnotesize]{caption}
    \usepackage[labelformat=simple, font=footnotesize]{subcaption}  
    \DeclareCaptionLabelSeparator{periodspace}{.\quad}
    \ifconf
    \else
    \fi
\else %
    \ifs %
    	\usepackage[small]{caption} 
    \else
    	\usepackage{caption}
    \fi
    \usepackage[labelformat=simple]{subcaption}
\fi

\usepackage{graphicx}
\usepackage{amssymb}
\usepackage{xurl}
\usepackage{xcolor} %

\usepackage{xspace}
\usepackage{balance}
\usepackage{booktabs}  %
	
\usepackage{array}
	\newcolumntype{C}[1]{>{\centering\let\newline\\\arraybackslash\hspace{0pt}}m{#1}}
\usepackage[inline,shortlabels]{enumitem} 
\usepackage{multirow}
\usepackage{tabularx}

\usepackage[capitalize]{cleveref}
\crefname{theorem}{Thm.}{Thms.}
\crefname{equation}{Eq.}{Eqs.}
\Crefname{equation}{Eq.}{Eqs.}
\crefname{figure}{Fig.}{Figs.}
\Crefname{figure}{Fig.}{Figs.}
\crefname{table}{Table}{Tables}
\Crefname{table}{Table}{Tables}
\crefformat{section}{\S#2#1#3}
\Crefformat{section}{\S#2#1#3}
\crefformat{subsection}{\S#2#1#3}
\Crefformat{subsection}{\S#2#1#3}

\ifcomm
	\newcommand{\mycomm}[3]{{\footnotesize{{\color{#2} \textbf{[#1: #3]}}}}}
     \newcommand{\Fmycomm}[3]{\footnote{{{\color{#2} \textbf{[#1: #3]}}} }}
\else
    \newcommand{\mycomm}[3]{}
    \newcommand{\Fmycomm}[3]{}
\fi

\newcommand{\new}[1]{#1}
\newcommand{\old}[1]{}

\ifs
	\usepackage[compact]{titlesec} %
		
	\newcommand{\ReduceVSpace}{\vspace{-13pt}}

    \usepackage[inline]{enumitem}
        \setlist{leftmargin=*} %
    \newcommand{\T}[1]{\par\vspace{2pt plus 1pt minus 1pt}\noindent\textbf{#1}} %

\else

	\newcommand{\ReduceVSpace}{}

    \usepackage[inline]{enumitem}
    \newcommand{\T}[1]{\par\smallskip\noindent\textbf{#1}} %
\fi

\newcommand{\be}{\begin{equation}}
\newcommand{\ee}{\end{equation}}

\newcommand{\para}[1]{\left( #1 \right)}        %

\newcommand{\unit}[1]{\,\mathrm{#1}} %

\newcommand{\vx}{\checkmark\kern-1.1ex\raisebox{.7ex}{\rotatebox[origin=c]{125}{--}}} %

\newcommand{\problong}{q\xspace}
\newcommand{\targdel}{target\_delay\xspace}
\newcommand{\ltcpu}{LTCP-U\xspace}
\newcommand{\swiftstar}{LSwift\xspace}
\newcommand{\ls}{\swiftstar}
\newcommand{\name}{MSwift\xspace}
\newcommand{\n}{NSCC\xspace}
\newcommand{\mn}{MNSCC\xspace}

\newcommand{\smartt}{SMaRTT\xspace}
\newcommand{\fastflow}{FastFlow\xspace}
\newcommand{\us}{\unit{\mu s}}

\newcommand{\appropto}{\mathrel{\vcenter{
  \offinterlineskip\halign{\hfil$##$\cr
    \propto\cr\noalign{\kern2pt}\sim\cr\noalign{\kern-2pt}}}}} %

\newcommand{\Tshort}{T_s}
\newcommand{\Tlong}{T_l}
\newcommand{\mm}{mm}

\newcommand{\newVar}[2]{\newcommand{#1}{\ensuremath{#2}\xspace}}
  \newVar{\server}{S}
  \newVar{\client}{C}
  \newVar{\rclient}{R_c}
  \newVar{\rserver}{R_s}
  
\providecommand{\vs}{{vs.}\xspace}
\providecommand{\ie}{{i.e.,}\xspace}
\providecommand{\eg}{{e.g.,}\xspace}

\providecommand{\etal}{{et al.}\xspace}

\begin{document}

\title{Congestion Control for Spraying \\
with Congested Paths}

\ifblind 
\else %
  \ifacm %
	\ifacmart
        \ifhyperref
            \newcommand{\aut}[2]{#1\texorpdfstring{$^{#2}$}{(#2)}}  %
        \else
            \newcommand{\aut}[2]{#1$^{#2}$}
        \fi
        \author{
          \aut{Grad Student}{1},
          \aut{Isaac Keslassy}{1,2}
        }%
        \affiliation{
        $^1$ \textit{Technion} \quad 
        $^2$ \textit{Somewhere else, Inc.} }
        \renewcommand{\shortauthors}{F. Author \textit{et al.}}     %
        \acmConference[CoNEXT'26]{ACM CoNEXT}{December 2026}{Seoul, South Korea}
        \settopmatter{printfolios=true,printacmref=false} %
           	  \setcopyright{none}
	\else %
    	\author{List of authors}
     \fi
  \else %
    \ifusenix %
      \author{
      {\rm Your N.\ Here}\\
      Your Institution
      \and
      {\rm Second Name}\\
      Second Institution
      } %
    \else %
        \ifhotnets %
            \author{TBD}
        \else 
            \ifieee %
               \Conf{
                    \ifhyperref
                        \newcommand{\aut}[2]{#1\texorpdfstring{$^{#2}$}{(#2)}}  %
                    \else
                        \newcommand{\aut}[2]{#1$^{#2}$}
                    \fi
                  \author{
                       \IEEEauthorblockN{%
                          \aut{Barak Gerstein}{1},
                          \aut{Mark Silberstein}{1,2},
                          \aut{Isaac Keslassy}{1,3}
                        }
                        \IEEEauthorblockA{
                              $^1$ \textit{Technion} \quad 
                              $^2$ \textit{NVIDIA} \quad 
                              $^3$ \textit{UC Berkeley}
                        }
                    }
                }
                \TR{\author{Grad Student and Isaac~Keslassy,~\IEEEmembership{Senior Member,~IEEE} \thanks{\textbf{[Add this 1st paragraph in 1st journal submission only:]} This paper was presented in part at IEEE Infocom '26, New York, May 2026. Additions to the conference version include new theorems, complete proofs that were previously omitted for space reasons, and additional simulation results.
        
                     G. Student and I. Keslassy are with the Department of Electrical and Computer Engineering, Technion, Israel (e-mails: \{grad@tx., isaac@\}.technion.ac.il).
                }}}
            \else %
                \author{TBD...}
            \fi
        \fi %
    \fi %
  \fi %
\fi %

\OnlyTR{\markboth{Technical Report TR16-01, Technion, Israel}{}}
\Journal{\markboth{}{}}%

\ifacmart
\else
	\maketitle
\fi

\ifacm %
    \sloppypar
\else 
    \ifhotnets
        \sloppypar
    \else
    \fi
\fi

\begin{abstract}
Packet spraying approaches are increasingly deployed in datacenter networks. However, their combination with existing congestion control algorithms (CCAs) may lead to poor QoS, especially when some of the paths are congested.

In this paper, we first model the throughput collapse of a wide array of CCAs when some of the paths are congested. We explain that since CCAs are typically designed for single-path routing, their estimation function focuses on the latest feedback and mishandles feedback that reflects multiple paths. We propose using a median feedback that is more robust to the varying signals that come with multiple paths. We introduce MSwift and MNSCC, which apply this median principle to Google's Swift and Ultra Ethernet's NSCC. We demonstrate that they can improve both CCAs, reaching better QoS both under congested paths and in uncongested networks.
\end{abstract} 

\ifacmart
	\maketitle
\else
\fi

\section{Introduction}
\label{sec:introduction}

RoCEv2 (RDMA over Converged Ethernet version 2) is used in many datacenter networks. It enables a wide set of applications, from AI training to Storage Area Networks (SANs). Unfortunately, many existing RoCE NIC implementations have historically been sensitive to reordering due to simple Go-Back-N (GBN) loss-recovery designs~\cite{irn}. 
Recent NICs  support  out-of-order delivery, but only for a subset of RDMA operations such as RDMA WRITE~\cite{nvidia_AR_whitepaper2}. However, other operations might still suffer from poor performance due to packet reordering.  
For example, RDMA READ operations send a multi-packet response where packets need to arrive in order, or else the late packets will be discarded and will have to be resent~\cite{irn}. As a result, RoCEv2 datacenters often rely on ECMP-based load-balancing techniques with a single path per flow. These techniques either prevent reordering entirely, as in Meta’s enhanced ECMP~\cite{meta}, or minimize it by using ECMP but allowing occasional re-pathing, as in FlowBender and Google’s PLB~\cite{flowbender,plb}.

However, with modern AI training workloads, such techniques are prone to flow collisions and yield poor performance~\cite{Stellar,nvidia_AR_whitepaper,li2025revisiting}. Thus, many companies increasingly rely on \textit{per-packet load balancing}, also called \textit{packet spraying}, to efficiently use their networks. For example, Alibaba implements host-based \textit{Oblivious Packet Spraying (OPS)}~\cite{RPS,LTCP,CAPS,Stellar}; NVIDIA's Spectrum-X~\cite{nvidia_AR_whitepaper2,nvidia_AR_roce} implements switch-based \textit{Adaptive Routing (AR)}~\cite{ARbook,nvidia_AR_whitepaper, broadcom_AR_whitepaper}; and the recent Ultra Ethernet Consortium (UEC) specification~\cite{uec} recommends host-based adaptive packet spraying algorithms such as REPS and STrack~\cite{reps, strack}. 

Therefore, modern networks must be efficient for the cases where per-packet spraying and per-flow ECMP variants must coexist. For example, NVMe-over-Fabric servers~\cite{nvme} heavily rely on RDMA READ operations which cannot run efficiently with per-packet spraying, whereas the NVIDIA NCCL collective library predominantly employs RDMA WRITE specifically to allow spraying~\cite{juniper-nccl,li2025revisiting}. Thus, most switch vendors support a concurrent use of both types of load balancing on the same network by marking for eligibility packets that can be sprayed (\eg Cisco~\cite{cisco-mix} and NVIDIA~\cite{nvidia_AR_whitepaper2}). In this case, the network congestion is highly non-uniform: some links are much more congested, as ECMP flows and sprayed packets are using these same links. %

The Congestion Control Algorithm (CCA) of the sprayed flows suffers in two ways from these congested links. The first problem is well known: the asymmetry causes  \textit{packet reordering}. This has been well studied in the past, especially with TCP~\cite{RPS, LTCP, fastRTO, reordering2007overview,network_coding_under_packet_reordering,Improving_TCP_performance_in_multipath_packet_forwarding_networks, Enhancing_TCP_performance_to_persistent_packet_reordering, tcp-dcr, tcp-rr, tcp-door}. When a packet traverses a high-latency path and arrives later than its predecessors, it can trigger duplicate ACKs and 
cause unnecessarily aggressive multiplicative decreases in the congestion window. 
This reordering problem can be mitigated by using Selective ACK (SACK) and delaying the time to perform congestion avoidance~\cite{LTCP}.

In this paper, we identify a second, previously unknown problem: \textit{a throughput collapse due to the recurrent congestion signals from the congested links}. Even if very few links are congested, 
packets crossing the congested links will repeatedly come back with congestion signals and cause the sender to aggressively reduce its rate.
This overreaction can lead to substantial under-utilization of available network capacity and to increased Collective Completion Times (CCTs) in AI training. 
The congested path can affect about any congestion signal, \eg (a)~a large delay in {delay-based CCAs} such as Google's Swift~\cite{swift}; 
(b)~an ECN-marked packet in  {ECN-based CCAs} such as DCTCP, which has been deployed in Meta's production datacenters~\cite{dctcp,judd,dctcp-meta}; or
(c)~a congestion signal in {In-band Network Telemetry (INT)-based CCAs} such as  Poseidon~\cite{poseidon}. 

We could think of several natural solutions to this throughput collapse. 
(i)~A first approach would be to allocate a different congestion window for each path, as in MPTCP or MP-RDMA~\cite{mprdma}. However, subflow-based load-balancing exhibits poor performance~\cite{ofan}. Packet spraying techniques employ a large number of paths for efficient load-balancing. The resulting  large number of windows would be hard to implement at scale, and many tiny windows would achieve poor results. Furthermore, in switch-based spraying techniques like AR, the hosts do not even know the path of the packets. 
(ii)~A second approach would be to employ packet spraying techniques like REPS that avoid sending over congested links. We show in evaluations that this does not help sufficiently (\cref{sec:eval}). 
(iii)~A third approach would be to separate ECMP flows from sprayed packets using traffic classes, and provide a fair scheduling in switches. In addition to its configuration complexity and interference with the existing QoS policies, we show that this approach still suffers from the throughput collapse, since sprayed packets keep encountering more congestion at shared links, and therefore hosts keep receiving recurrent congestion signals (\cref{sec:eval}).

\T{Contributions.} 
In this paper, we make three main contributions. First, we introduce a fundamental model of throughput collapse induced by the joint use of packet spraying and ECMP. We assume a simple scenario where packets go through a congested path with probability $\problong$, while other paths stay uncongested. We demonstrate a general rule: \textit{Throughput collapses as $\Theta(1/\sqrt{\problong})$}.
We prove that it applies to five different CCAs: (1)~TCP (New Reno); (2)~DCTCP; (3)~Swift; 
(4)~\smartt/\fastflow~\cite{fastflow}, the basis of the \n CCA in the UEC specification~\cite{uec} that is designed for packet spraying; 
and (5)~a reordering-resilient Swift variant. This occurs even though these CCAs use different congestion signals in different ways.

Second, to solve the throughput collapse, we propose to adapt CCAs by using a general \textit{median-based framework}. Instead of responding only to the latest congestion signal, as we would when signals come from a single path, this simple approach tracks recent congestion signals and reacts based on their median value. Thus, this approach 
remains robust against varying per-packet multi-path signals, while following the intended goals of the CCA. We explain how this framework can also be fine-tuned using a Nyquist perspective. We illustrate the framework with two CCAs: (i)~based on Google's Swift, we implement  \textit{\name}, where we apply our median-based framework to a reordering-resilient Swift baseline; and (ii)~based on UEC's \n, we introduce \textit{\mn}. Extensive evaluations using 3-layer fat-tree topologies show that \name improves the sprayed-traffic CCT inflation in the presence of ECMP flows by up to $8.6\times$ using REPS. It outperforms both Swift and \n in all 3 compared spraying algorithms: OPS, AR and REPS.  In addition, using fair DRR scheduling at switches exhibits similar results.

Our third contribution is to show that this new CCA framework does not only help with permanent ECMP congestion, but also with temporary congestion without ECMP flows. We simulate training a Llama 70B model using HSDP, and find that \name outperforms all other algorithms on all spraying techniques. Its CCT inflation with REPS is $1.8\times$ smaller than a reordering-resilient Swift and $2.3\times$ smaller than \n.

\section[The Theta(1/sqrt(q)) CCA Multi-Path Rule]{The $\Theta(1/\sqrt{q})$ CCA Multi-Path Rule} %
\label{sec:model}

\begin{figure}[!t] %
\centering
	\includegraphics[width= 0.35 \textwidth]{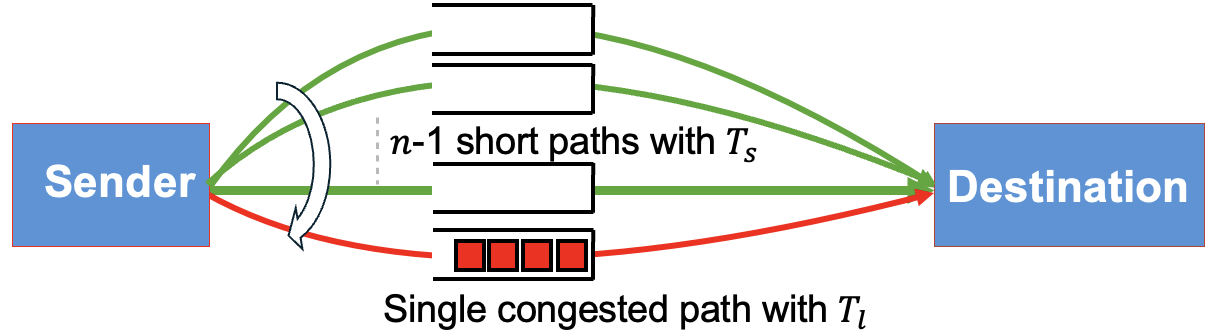}
	\caption{Scenario with single congested path.
		}
	\label{fig:model settings}
\ReduceVSpace %
\end{figure}

In this section, we start by presenting a theoretical analysis that gives us a more rigorous understanding of the impact of a few congested paths on packet-sprayed flows using existing CCAs. We show that it causes the throughput collapse of these flows. This result leads us to propose \name in \cref{sec:mswift}.

\subsection{CCAs}
We model and demonstrate throughput collapse in five core CCAs that can be used in datacenters and rely on different congestion signals.

\T{(1) TCP.} We start with the basic TCP New Reno, as it is a fundamental inspiration for later CCAs. 

\T{(2) DCTCP,} %
which
uses both 
loss signals and ECN marking. 

\T{(3) Swift.} Google's Swift~\cite{swift}
strongly relies on packet delays, as well as on loss signals.

The above 3 CCAs are reordering-sensitive.
We also demonstrate that the same throughput collapse applies to 
two additional reordering-insensitive CCAs: 

\T{(4)~\smartt.} \smartt~\cite{fastflow}, also known as \fastflow, is the initial default CCA for REPS. It is important as it forms the basis for NSCC, the mandatory CCA in the UEC specification~\cite{uec}. It 
relies on
{delay} and {ECN} congestion signals.

\T{(5)~Reordering-resilient Swift.} This variant is adopted to neutralize the impact of reordering and show that throughput collapse happens despite it. 

\subsection{Model}
\cref{fig:model settings} illustrates our theoretical model. We consider a single flow that sprays packets 
across $n$ paths. We assume that a single path out of $n$ always experiences congestion.  Therefore, the probability of a random packet to experience congestion is $q=1/n$. We also assume a constant-delay congestion to make the model tractable. 
We denote the larger RTT of the congested path as $\Tlong$ and the shorter RTT of the uncongested paths as $\Tshort$.

\subsection{Results}

Our analysis reveals that if the flow uses TCP, DCTCP, Swift, \smartt, or a reordering-resilient Swift as CCAs, it will experience a throughput collapse that follows a common scaling rule: %
$$throughput = \Theta\para{\frac{1}{\sqrt{q}}}.$$

\begin{table}
\caption{CCA multi-path throughput models when congestion occurs with probability $q$ and leads to a path delay of $\Tlong$ rather than $\Tshort$. MSS stands for maximum segment size, $\mm\equiv max\_mdf$ and $ai$ are Swift parameters.
The throughput of all these CCAs falls like $\Theta(1/\sqrt{q})$, experiencing a sharp drop even with little congestion. }
\label{tab:throughput-models}
\centering
\renewcommand{\tabularxcolumn}[1]{>{\centering\arraybackslash}m{#1}}
\begin{tabularx}{0.99\columnwidth} {c >{\centering\arraybackslash}X c >{\centering\arraybackslash}X}
	\toprule
	\textbf{Thm.} & \textbf{CCA} & \textbf{Throughput Model} & \textbf{Collapse Reason} \\
	\midrule
	\cref{tcp theorem} & TCP     & $\frac{MSS}{\Tshort}\frac{1.22}{\sqrt{\problong}}$ & Packet Reordering \\
	\Cref{dctcp theorem} & {DCTCP (upper bound)}   & $\frac{MSS}{\Tshort}\frac{1.22}{\sqrt{\problong}}$ & Packet Reordering \\
	\Cref{original swift theorem} & Swift   & 
	$\frac{MSS}{\Tshort}\frac{\sqrt{\para{1/\mm-1/2}\cdot ai}}
	{\sqrt{\problong}}$ & Packet Reordering \\
	\Cref{FastFlow theorem} & \smartt  & {$\Theta(1/\sqrt{q})$} & Delay + ECN Signals  \\
	\Cref{reordering-resilient Swift theorem} &
	{Reordering-resilient Swift} & {$\Theta(1/\sqrt{q})$} & Delay Signal \\
	\bottomrule
\end{tabularx}
\end{table}

Given the general model above, we make additional simplifying assumptions that mostly follow the standard Mathis model assumptions~\cite{mathis}. For example, we assume that paths are taken in a round-robin order rather than at random, and that the flow avoids retransmission timeouts. We also make CCA-specific assumptions. For example, for DCTCP, we distinguish whether the long path triggers congestion or not. All  assumptions and proofs are detailed in \cref{sec:proofs}. Under these assumptions, as \cref{tab:throughput-models} summarizes, we obtain the following theorems:

\begin{theorem}[TCP]\label{tcp theorem}
The throughput of a single TCP flow is
\begin{equation}
	throughput = \frac{MSS}{\Tshort}\frac{1.22}{\sqrt{\problong}}
	\label{eq:TCP 2 RTTs formulas}
\end{equation}
\end{theorem}

\begin{theorem}[DCTCP]
\noindent{(i)}~If a long-path packet doesn't get ECN-marked, the throughput of a single DCTCP flow is
\begin{equation}
	throughput = \frac{MSS}{\Tshort}\frac{1.22}{\sqrt{\problong}}
	\label{eq:DCTCP unmarked}
\end{equation}
\noindent{(ii)}~Else, \cref{eq:DCTCP unmarked} is a \emph{throughput upper bound}, so 
\begin{equation}
    throughput = O(1/\sqrt{q})
	\label{eq:DCTCP marked}
\end{equation}
\label{dctcp theorem}
\end{theorem}

\begin{theorem}[Swift]
The throughput of a single Swift flow is
\begin{equation}
	throughput = \frac{MSS}{\Tshort}\frac{\sqrt{\para{\frac{1}{\mm}-\frac{1}{2}}\cdot ai}}{\sqrt{\problong}} 
	\label{eq:throughput max mdf}
\end{equation}
\label{original swift theorem}
\end{theorem}

\begin{theorem}[\smartt]
The throughput of a single \smartt flow is
\begin{equation}
	throughput = \Theta(1/\sqrt{q})
	\label{eq:throughput FastFlow}
\end{equation}
\label{FastFlow theorem}
\end{theorem}

\begin{theorem}[Reordering-resilient Swift]
The throughput of a single reordering-resilient Swift flow is 
\begin{equation}
    throughput = \Theta(1/\sqrt{q})
	\label{eq:throughput reordering resilient swift}
\end{equation}
\label{reordering-resilient Swift theorem}
\end{theorem}

\begin{figure}
\centering
\includegraphics[width= 0.7  \columnwidth]{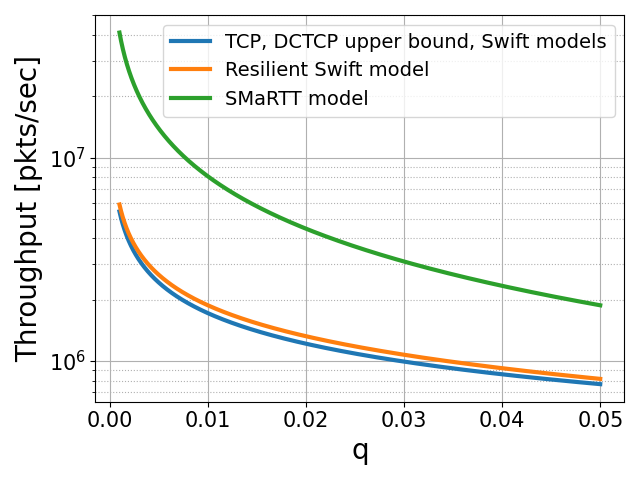}
\caption{Model comparison using the theoretical results in \cref{tab:throughput-models}.
}
\label{fig:comparison}
\ReduceVSpace %
\end{figure}

\T{Illustration of throughput collapse.} \cref{fig:comparison} illustrates this throughput collapse as the proportion $q$ of congested paths  increases.
\textit{The main takeaway is that the impact of even a few congested paths is rather dramatic. Having 2\% of congested paths rather than 1\% decreases the throughput by a factor of $\sqrt{2}\approx 1.4 \times$, a large impact in datacenters, with an even worse effect on CCTs.}
\Cref{fig:comparison} also shows that reordering-resilient CCAs yield a higher throughput, but are still subject to the throughput drop.

\begin{figure}
\centering
\includegraphics[width=0.35 \textwidth]{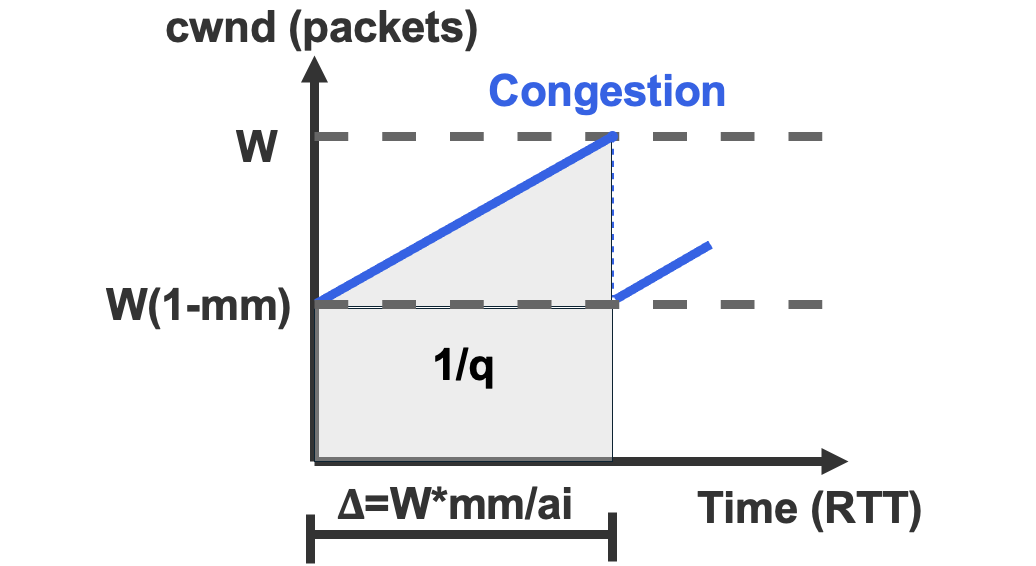}
\caption{AIMD dynamics for Swift, providing some intuition for why the grey area of $1/q$ is roughly proportional to $W \times W$. 
} 
\label{fig:multiplicative decrease}
	\ReduceVSpace %
\end{figure}

\T{Proof intuition.}  \cref{fig:multiplicative decrease} illustrates the proof intuition for Swift. It reflects a general intuition behind this rule for algorithms employing an \textit{additive increase and multiplicative decrease} (AIMD) policy for their congestion window $W$. Note that the intuition does not equally apply to all algorithms, \eg \smartt does not only use AIMD and its proof is more complex.
As \cref{fig:multiplicative decrease} illustrates using the well-known sawtooth pattern, intuitively: (i)~the flow sends a number of packets proportional to the congestion window (cwnd) size $W$ at each RTT (\eg the sawtooth will oscillate between $W/2$ and $W$), and (ii)~in congestion avoidance, at each RTT, cwnd increases by a constant, so to grow from $W/2$ to $W$ will take a number of RTTs proportional to $W$. Combining both remarks yields that a flow sends a number of packets proportional to $W^2$ between congestion events. If there is a congestion event every $n=1/q$ packets, then $W^2 \propto 1/q$ and $W \propto 1/\sqrt{q}$.  

The rule is closely related to the Mathis $\Theta(1/\sqrt{p})$ throughput model for TCP given a loss rate of $p$~\cite{mathis}. The reason is that in TCP, both losses and late packets are detected through the same mechanism of three duplicate ACKs. 
However, both rules are not identical. If path delays in the congested path were replaced by drops to apply the Mathis model, the $\Theta(1/\sqrt{p})$ model would not hold for several of our CCAs. In fact, \smartt and reordering-resilient Swift sometimes do not even reduce their congestion windows upon packet drop.

\section{\name}\label{sec:mswift}

\subsection{Median framework for CCAs}

\T{Design goals.} 
We seek a new robust framework that can modify CCAs to let them handle packet-sprayed traffic that encounters a few congested links, without causing throughput collapse. The framework should (i)~be simple; (ii)~improve performance for datacenter traffic workloads such as permutation traffic; 
(iii)~apply to most advanced CCAs; and (iv)~maintain compatibility with single-path deployments without performance degradation. 

\T{Key idea: using the median.} We want to avoid drastic throughput reductions due to intermittent congestion signals originating from different paths. To do so, we propose using \textit{the median value} of recently received congestion signals.
This median-based approach has three key properties: (i)~The median is well known in robust statistics as a robust measure of central tendency~\cite{hampel2001robust}. Thus, the median of the latest congestion signals intuitively attempts to capture the overall state tendency of the network 
by generating a coalesced signal that is insensitive to outliers. (ii)~Using a median, rather than another percentile (\eg min or max of recent signals), better fits the initial goals of the CCA designer. For example, the Swift CCA attempts to maintain the typical packet delay close to some target. Adopting a median follows this logic. Adopting a more aggressive approach would deviate from the initial CCA goal. (iii)~Computing an approximate median is feasible at high speeds, as we explain below.  

Let's prove that a median-based CCA can avoid throughput collapse. Assume a standard ACK-based CCA that uses the median framework with a median window of at least 3 ACKs and with pacing. Then we get:

\begin{theorem}[Median-based CCA]
\label{thm:median}
An infinite flow that uses the above median-based CCA avoids the throughput collapse whenever $q<1/3$.
\end{theorem}
The reason for assuming an infinite flow  is that in an adversarial case, a flow may get back all of its ACKs for the uncongested packets, stop sending, then obtain much later the ACKs for its congested packets. Thus, a median-based CCA would decide that the network is congested. (In practice, we could ignore congestion signals that are older than 1-2 RTTs to avoid this corner case.)  However, in a typical case where the flow keeps sending traffic, a median-based CCA can make sure that the majority of uncongested ACKs prevails.

\subsection{\name and \mn}

The median-based principle can be broadly applied across a variety of congestion signals and algorithms, \eg \textit{delay measurements} in Swift, \textit{ECN marks} in DCTCP, and \textit{MPD values} from in-network telemetry in Poseidon. Since applying it to all these CCAs is beyond the scope of this single paper, we show how this median-based framework can be applied to two leading datacenter CCAs: Google's Swift and UEC's \n.

\T{\name.}
First, we want to apply to apply the median-based framework to a reordering-resilient version of Swift. We use the core idea of LTCP~\cite{LTCP} %
(with the \ltcpu variant) to design and implement \textit{\swiftstar}, a reordering-resilient Swift baseline. We choose LTCP because of its specific design for datacenter packet spraying and its SACK-based operation, enabling seamless Swift integration. We adjust the multiplicative window reduction trigger from two to five successive delayed packets, to allow more reordering in congested datacenters.

Then, we propose and implement \textit{\name}, which builds upon our reordering-resilient baseline {\swiftstar}, and applies our proposed median-based approach. 
For its decisions, \textit{\name uses the median delay of recent ACKs} rather than the delay of the latest ACK. 

\T{\mn.} \n is designed for spraying and does not retransmit packets upon receiving a duplicate ACK. Thus, we do not need to make it reordering-resilient. We can directly apply the median framework to the delay signals in \n and call the resulting algorithm \textit{\mn}. 

\T{Lightweight implementation.} Adding \name to  \swiftstar and \mn to \n takes less than a total of 200 lines of code in htsim. While a median computation is harder in hardware, a pseudo-median approximation is feasible at high speeds. It can be implemented using cascaded max/min comparators and pipelined, albeit at the cost of potentially significant distances from the true median if the congestion signals can span a wide range.

\subsection{Nyquist rate}

The median-based framework has a single parameter that needs to be tuned: the \textit{history size} $H$, \ie the number of latest congestion signals used to compute the median. For example, in \name and \mn, we want to compute the median of the delays obtained using the last $H$ ACKs.

$H$ is the result of a tradeoff. On the one hand, a high $H$ enables stable decisions based on the median of a long history. On the other, $H$ needs to be small enough to capture the frequent variations of the network. If the network varies once per $RTT$, $H$ should not equal several RTTs. Likewise, when a congestion starts, a high $H$ will delay the answer.

When we keep taking the median signal of the last $H$ packets, we obtain a transformed signal that is only independent every $H$ packets, \ie we intuitively cut the signal into independent ranges of size $H$. This is related to 
sampling the signal once every $H$ packets. The network state changes as the cwnd values of all flows change. How to set $H$  to accurately capture the variations of cwnd in this sampling? This sampling should act as a low-pass filter, aliasing out the high-frequency ACK clocking behavior and capturing the macroscopic fluid-model behaviors of the flows.

To do so, we rely on the \textit{Nyquist rate}. Assuming that the cwnd $W$ of each flow in the network grows by about 1 every $K$ packets, we want to set $H=K/2$, \ie sample every $K/2$ packets to accurately track this growth. Since the median needs at least one packet, we get:
\be H=\max\para{K/2,1}.
\ee 
Of course, we do not claim that the CCA signal is bandpass-limited, nor that we can apply the Nyquist–Shannon sampling theorem. However, this will provide us with a rule of thumb for establishing $H$.

\begin{itemize}
    \item In Swift, cwnd grows by $ai=1$ every RTT, \ie roughly every cwnd of $W$ packets. Thus,
    \be H_{\text{\name}}=\max\para{W/2,1}.
    \ee
    
    \item In \n, cwnd grows by about $1$ either every 8 packets or every RTT, whichever comes first. We approximate it as $K=\min\para{W,8}$, yielding:
    \be H_{\text{\mn}}=\max\para{\min\para{W/2,4},1}.
    \ee
\end{itemize}

We can see that more aggressive CCAs like \n will use a lower $H$, getting closer to the original single-ACK signal ($H=1$), thus their improvement is expected to be smaller. This will be confirmed in the evaluations.

\section{Evaluations}
\label{sec:eval}

\subsection{Settings}

\T{Simulator.} All of our evaluations are performed using (and extending) the htsim packet-level network simulator~\cite{htsim}, using an htsim fork~\cite{htsim-fork} to get an OPS implementation for Swift, and the official Ultra Ethernet repository~\cite{htsim-uec} to obtain the reference implementation of NSCC.\footnote{During our use of htsim, we corrected its Swift implementation to match the Swift mechanisms in the paper, especially (1)~the SACK mechanism, and (2)~the target delay calculation.
We also corrected the packet-spraying method used in the fork.}

\T{Topology.} We use a standard 3-layer, non-blocking fat-tree topology with 128 nodes, 800\,Gbps links, $0.5\us$ link latency, $800\unit{KB}$ per-port buffers, and RoCEv2-sized $4\unit{KB}$ packets~\cite{ofan}. Switches use standard droptail FIFO queues with ECN and without trimming.

\T{CCAs.} We compare five CCAs: (i)~Swift, (ii)~\swiftstar, (iii)~\name, (iv)~\n and (v)~\mn. For all CCA algorithms, we use the default parameter values adjusted for our network settings. Swift targets a low queueing delay, which can be crucial in datacenters for short flows. For an apples-to-apples comparison, we set \n's target queueing delay to equal that of Swift (about $1\us$), and set a switch ECN threshold of $40\unit{KB}$. To mimic the behavior of RDMA's persistent Queue Pairs (QPs) in datacenters, we set each initial cwnd to its ideal value, \ie start sending at high rate rather than waiting many RTTs to converge.

\T{LB algorithms.} We use three packet-spraying algorithms with their default htsim implementations: (i)~OPS, where the source host picks a random flow label for each packet; (ii)~AR, where each switch forwards packets to the next-hop port with the shortest quantized queue; and (iii)~REPS, where the source host sets random flow labels at first, and upon receiving ACKs, keeps re-using flow labels without ECN congestion. We use ECMP for background elephant flows.

\T{Baseline workload: permutation with elephants.} Our baseline workload consists of a mix of a sprayed permutation traffic with background elephant ECMP flows, reflecting the goal of this paper of studying the impact of such ECMP flows on the performance of sprayed traffic. To get the background traffic, we pick four random hosts and draw a random permutation among them, resulting in four ECMP long-lived flows. Then, we draw another random permutation among the remaining 124 hosts. Each of the 124 flows uses packet spraying to send $8\unit{MB}$. The ECMP and sprayed flows use the same CCA to avoid any fairness issues: making the sprayed flows more aggressive also makes the ECMP ones more aggressive. Note that the 4 ECMP flows only form about 3\% of the flows, but each ECMP flow uses up to 6 links, so the probability for a random sprayed packet to hit at least one of them is larger (about 0.09).

We later measure the sensitivity of this baseline workload by varying the number of hosts, the message size, the number of ECMP flows and the link failure rate. 

We also study a scenario where each switch implements DRR fair scheduling. Sprayed flows and ECMP flows are marked into two classes of traffic, and each class of traffic gets its own FIFO queue that is served by the DRR scheduler, with equal 50/50 rates and DRR $4\unit{KB}$ quantum size.

\T{Additional workloads.} We study three additional workloads: \\
\noindent (i)~A \textit{permutation} traffic without elephants, using all 128 hosts and $8\unit{MB}$ messages. This workload helps to evaluate whether the median framework also helps when there are no elephants. \\
\noindent (ii)~An \textit{AI training} workload. This workload helps evaluate the CCAs with a more realistic scenario. We simulate training  Llama-3 70B with \textit{Hybrid Sharded Data Parallel (HSDP)}~\cite{zhao2023pytorch} (FSDP2 2D) on our 128-GPU cluster organized as 16 servers of 8 GPUs, and we assume random server placement. HSDP traffic is implemented as a hierarchical ring. Intra-server communications are performed in the high-bandwidth domain, while our fat-tree network implements inter-server communications using 8 parallel rings: each logical GPU $i$ is connected to logical GPU $i+8 \pmod{128}$. We simulate one ring step of the backward pass of HSDP, translating to 3,344 packets per flow (assuming FP8 precision and 80 layers).  \\
\noindent (iii)~\textit{Incast} traffic without elephants, where 32 random hosts send $8\unit{MB}$ flows to a common destination host. The goal is to evaluate how much the median framework hurts in an incast scenario that can need quick cwnd variations. %

\T{Metrics.} We collect (i)~the Flow Completion Time (FCT) of each flow, and  
(ii)~the Collective Completion Time (CCT), which is the worst-case FCT across all flows. For this CCT to be meaningful, we show its increase above a zero-queueing lower bound~\cite{ofan}. The error bars indicate the standard error of the mean across runs.

\subsection{Baseline workload}

\begin{figure}
\centering
\includegraphics[width= 0.6 \columnwidth]{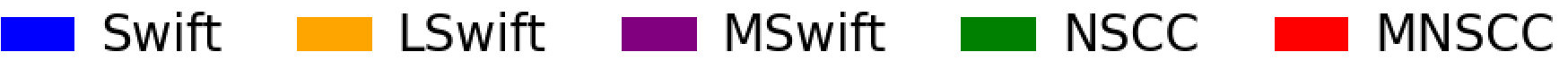}
\includegraphics[width= \columnwidth]{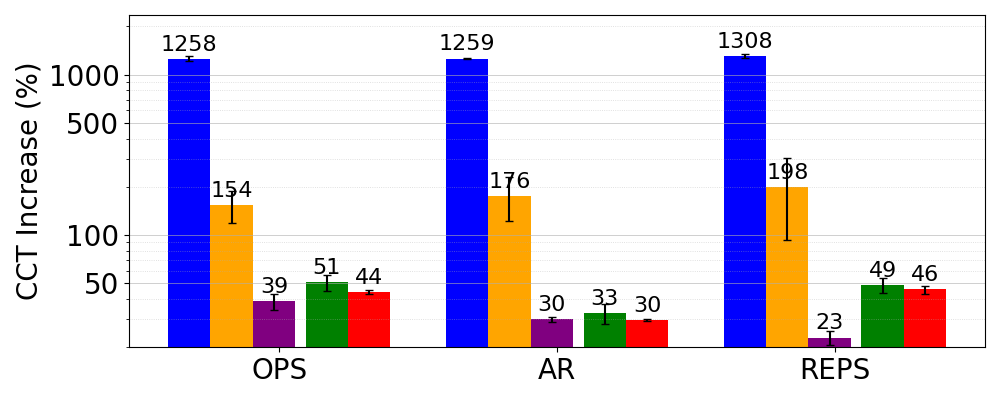}
\caption{%
Baseline workload: sprayed permutation flows with 4 ECMP flows.} 
\label{fig:1}
\ReduceVSpace %
\end{figure} 

\Cref{fig:1} illustrates the CCT increase in the baseline workload. It compares the five CCAs using three LB algorithms. First, it shows the particularly poor performance of the Swift CCA when using packet spraying, reaching some $13\times$ higher CCTs. This reflects Swift's aggressive cwnd reduction upon SACK holes following out-of-order arrivals. Thus, we exclude Swift from the next evaluations. \ls is designed to deal with this out-of-order issue, but still obtains a poor performance that illustrates the throughput collapse demonstrated in this paper. \name is designed to deal with this throughput collapse, and obtains the strongest performance across all algorithms (with REPS). When compared to \ls, it reduces the CCT inflation by up to $8.6\times$. 
\n also obtains good performance, as it is designed for packet spraying, and its delay averaging methods reduce the impact of elephant flows. \mn improves it a bit, but does not reach the performance of \name. Finally, we see that REPS reaches the best performance with \name, as it is designed to learn from wrong paths, but interestingly it is not the best LB algorithm for all CCAs. For example, even though REPS was designed together with \smartt, the inspiration for \n, it is not the best LB algorithm for \n, which performs better with AR.

\begin{figure}%
\centering
\includegraphics[width= 0.5 \columnwidth]{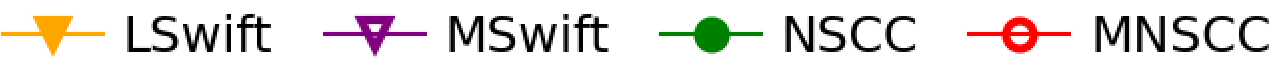}\par
\begin{subfigure}{0.32\linewidth}
\includegraphics[width=\textwidth]{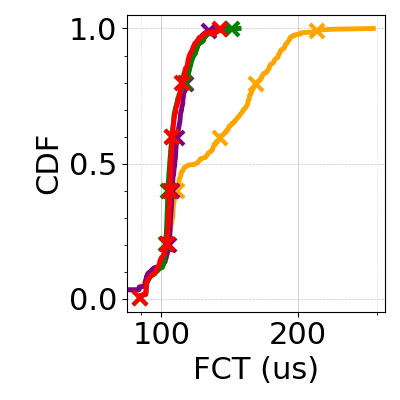}
  \caption{OPS}
  \label{fig:1a}
\end{subfigure}
\hfill
\begin{subfigure}{0.32\linewidth}
\includegraphics[width=\textwidth]{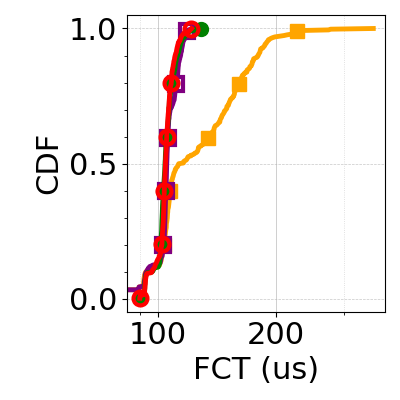}
  \caption{AR}
  \label{fig:1b}
\end{subfigure}
\hfill
\begin{subfigure}{0.32\linewidth}
\includegraphics[width=\textwidth]{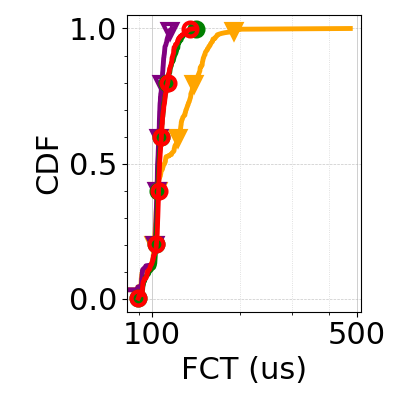}
  \caption{REPS}
  \label{fig:1c}
\end{subfigure}
\caption{CDF of FCT with baseline.}
\label{fig:1fct}
\ReduceVSpace %
\end{figure}

\Cref{fig:1fct} plots the CDF of the FCT using each CCA and each spraying approach. \ls clearly has the worst performance in all spraying schemes, and in particular it has a poor tail with REPS (top right of \cref{fig:1c}). \name performs similarly to \n and \mn with OPS and AR. However, \name's tail behaves better with REPS (top left), leading to its lower CCT (in \Cref{fig:1}).

\subsection{Permutation workloads}

\begin{figure}
\centering
\includegraphics[width= 0.5 \columnwidth]{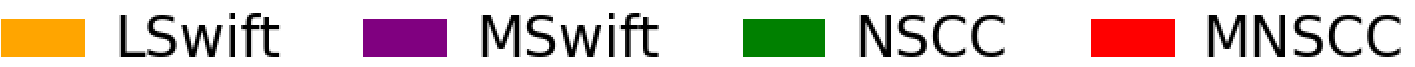}
\includegraphics[width= \columnwidth]{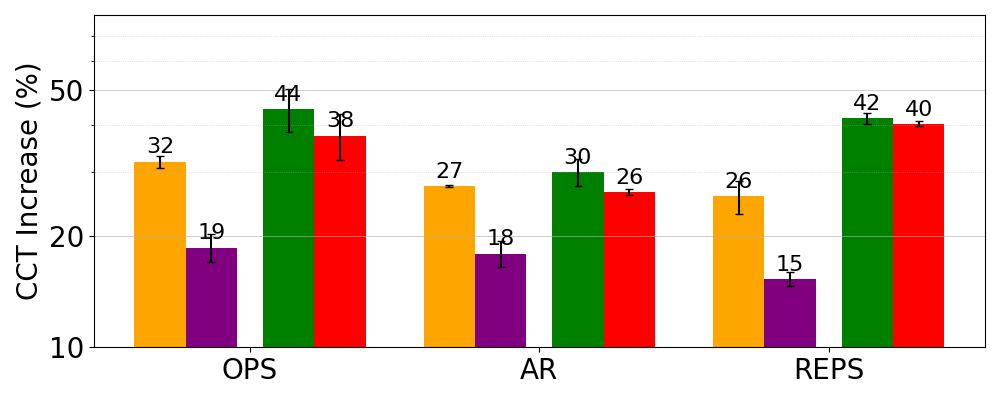}
\caption{Sprayed permutation workload (without ECMP flows).}
\label{fig:3}
\ReduceVSpace %
\end{figure} 

We now want to study how the median approaches fare in permutation workloads without ECMP flows, even though they were not designed for such workloads.

\Cref{fig:3} illustrates the CCT increase for a simple random permutation workload with no ECMP flows. All algorithms perform better when the ECMP flows are removed (compared to \Cref{fig:1}), and in particular \ls is much improved and becomes better than \n. It is significant that even though there are no ECMP flows, \name always improves upon \ls (by up to $1.7\times$), and in fact obtains the best performance with REPS. This shows that its fundamental mechanism can also help when congestion is temporary and not permanent. \mn slightly improves upon \n, yet \name still outperforms both.

\begin{figure}
\centering
\includegraphics[width= 0.5 \columnwidth]{fig/l2.png}
\includegraphics[width= \columnwidth]{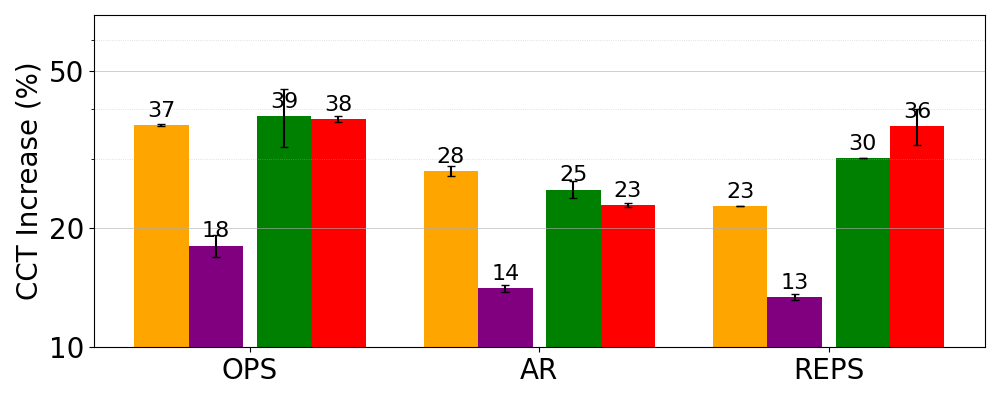}
\caption{AI training workload for Llama 70B model with HSDP.}
\label{fig:12}
\ReduceVSpace %
\end{figure} 

\Cref{fig:12} shows the results of the AI training workload, simulating an HSDP training of the Llama 70B model. \name outperforms again all other algorithms given all spraying techniques. Its CCT inflation with REPS is $1.8\times$ smaller than \ls and $2.3\times$ smaller than \n. Unlike \name,  \mn does not help much in this scenario without ECMP traffic that it is not designed for.

\subsection{Sensitivity to parameters}

We now study the sensitivity of the baseline workload to its various parameters.

\begin{figure}
\centering
\includegraphics[width= 0.5 \columnwidth]{fig/l2.png}
\includegraphics[width= \columnwidth]{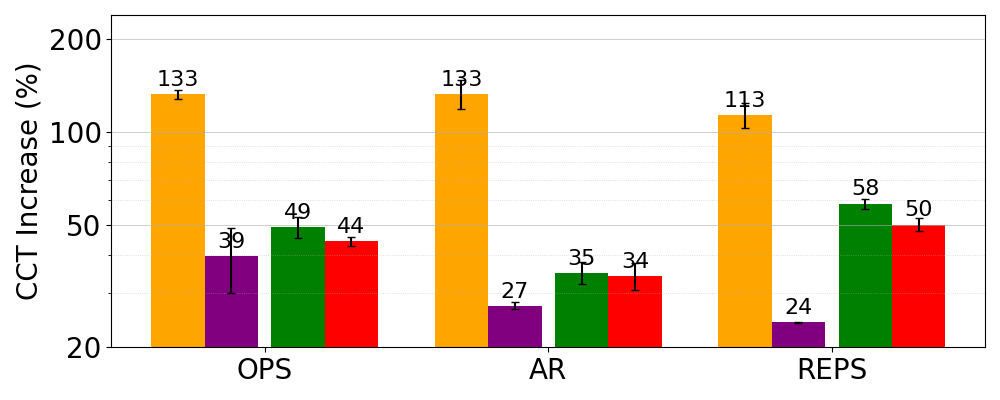}
\caption{Scaled baseline workload with 250 hosts.}
\label{fig:7}
\ReduceVSpace %
\end{figure} 

\T{Network size.} \Cref{fig:7} shows the results of increasing the three-level fat-tree network size from 128 to 250 nodes while keeping four elephants (and $250-4=246$ sprayed flows). The scaled network yields similar results. \name again outperforms all other approaches. \mn keeps slightly improving \n.

\begin{figure}
\centering
\includegraphics[width= 0.5 \columnwidth]{fig/l2.png}
\includegraphics[width= \columnwidth]{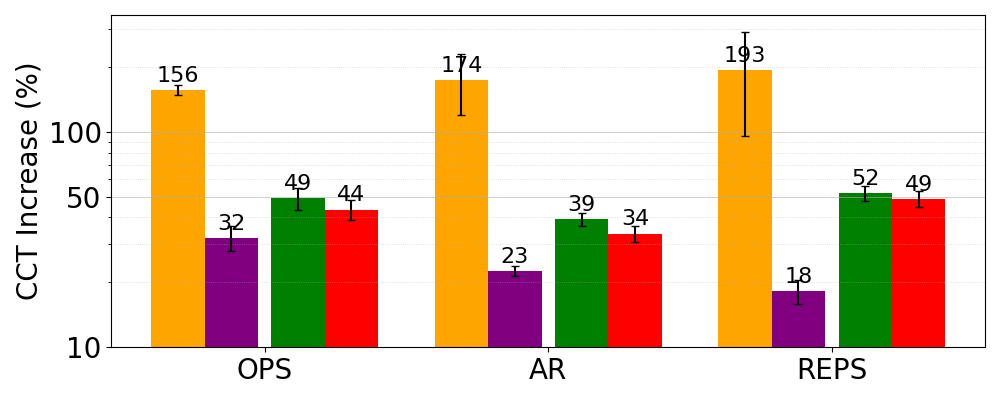}
\caption{Baseline workload with a doubled message size.}
\label{fig:8}
\ReduceVSpace %
\end{figure} 

\T{Message size.} \Cref{fig:8} illustrates the evaluation results with a doubled message size of $16\unit{MB}$. \name obtains even stronger results as there is more time for the CCA mechanisms to make a difference, with up to $10.7\times$ reduction in the CCA inflation \vs \ls, while both \ls and \n tend to get worse.

\begin{figure}
\centering
\includegraphics[width= 0.5 \columnwidth]{fig/l2.png}
\includegraphics[width= \columnwidth]{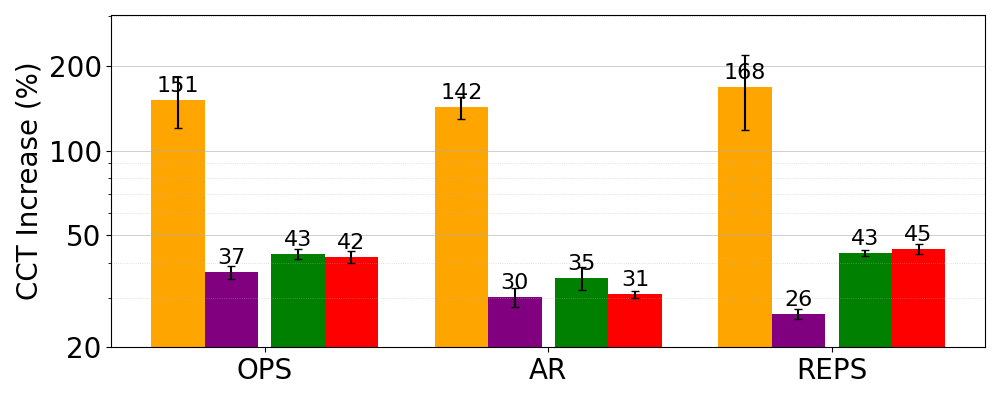}
\caption{Baseline workload with a doubled number of ECMP flows.}
\label{fig:9}
\ReduceVSpace %
\end{figure} 

\T{ECMP flows.} \Cref{fig:9} displays the results with a doubled number of ECMP flows in the baseline workload, \ie 8 instead of 4 (and $128-8=120$ sprayed flows). Interestingly, results are similar and not much worse than with 4 ECMP flows, maybe because the CCT focuses on the worst case and even though more flows experience congestion, the worst-case congestion does not change much. 

\begin{figure}
\centering
\includegraphics[width= 0.5 \columnwidth]{fig/l2.png}
\includegraphics[width= \columnwidth]{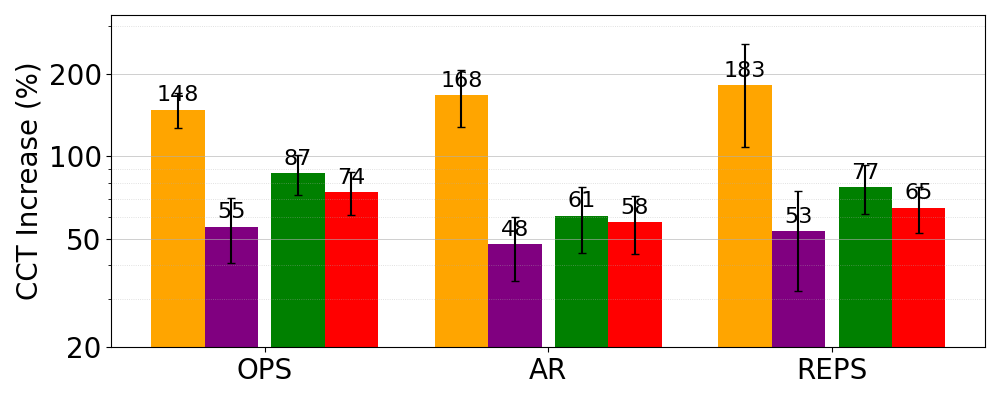}
\caption{Baseline workload with 1\% link failure rate.}
\label{fig:10}
\ReduceVSpace %
\end{figure} 

\T{Random failures.} \Cref{fig:10} shows the results of failing each inter-switch link with a 0.01 probability. Results are a bit noisier, because each run uses a new topology with different failed links, but we use the same topology for all algorithms so the relative results are more reliable. As expected, CCT is higher as there is no full network capacity anymore. However, \name still keeps outperforming with the three spraying algorithms.

\begin{figure}
\centering
\includegraphics[width= 0.5 \columnwidth]{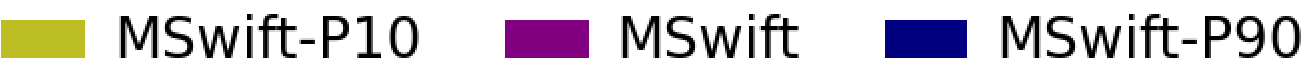}
\includegraphics[width= \columnwidth]{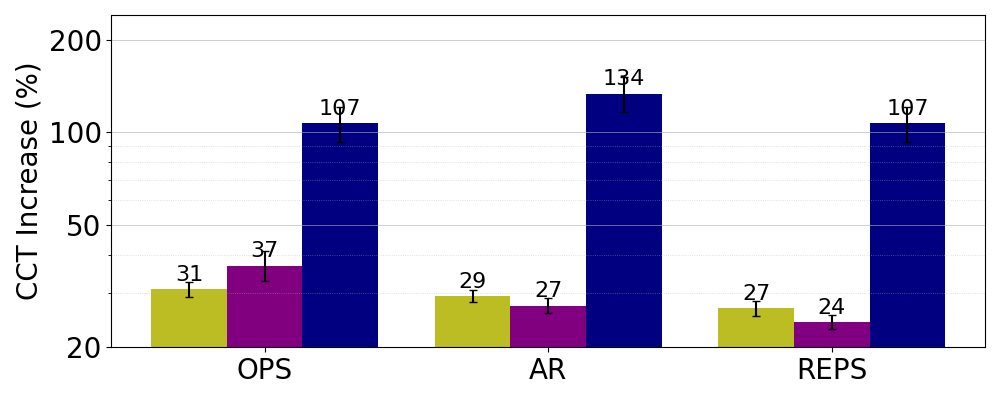}
\caption{Baseline workload with \name variants.}
\label{fig:11}
\ReduceVSpace %
\end{figure} 

\T{\name variants.} \Cref{fig:11} displays the results of using \name variants that are not medians (\ie 50th percentile): either an aggressive MSwift-P10 that only considers the 10th percentile of delays, \ie the lowest delays, and therefore does not back off easily; or a conservative MSwift-P90. We do not intend to adopt either, as they are incompatible with the Swift queueing delay target. However, we want to see if \name's CCT worsens much because \name is not aggressive enough. We can see that with OPS, the more aggressive MSwift-P10 outperforms \name, but in the other cases being overly aggressive or conservative does not pay off.

\subsection{Additional scenarios}

We now want to check whether the performance of median-based algorithms does not degrade too much in scenarios that the median framework was not designed for.

\begin{figure}
\centering
\includegraphics[width= 0.5 \columnwidth]{fig/l2.png}
\includegraphics[width= \columnwidth]{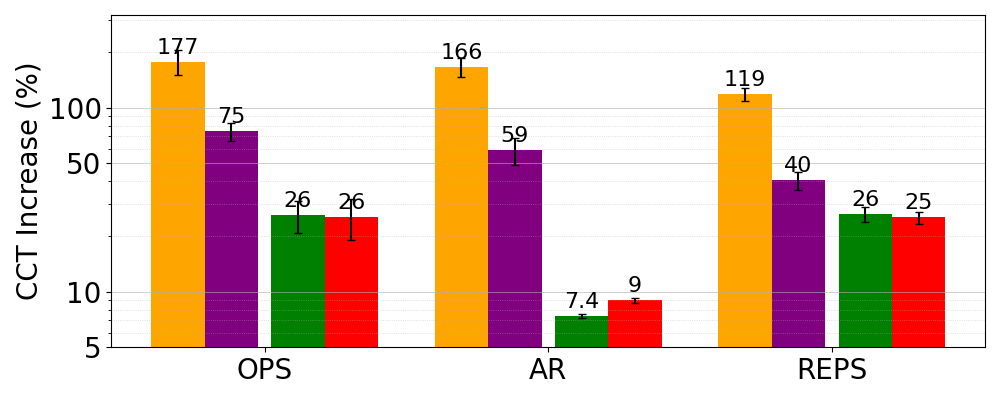}
\caption{Baseline workload with fair switch scheduling}
\label{fig:2}
\ReduceVSpace %
\end{figure} 

\T{Fair scheduling.} \Cref{fig:2} shows the behavior of the baseline workload under a fair DRR switch scheduling between sprayed and ECMP packets. We learn two lessons: First, Swift variants tend to get worse, while \n variants tend to get better. This may be related to the different effects of fair scheduling on the delay and ECN congestion signals. Second, \name still outperforms \ls, but \mn does not help with \n anymore.

\begin{figure}
\centering
\includegraphics[width= 0.5 \columnwidth]{fig/l2.png}
\includegraphics[width= \columnwidth]{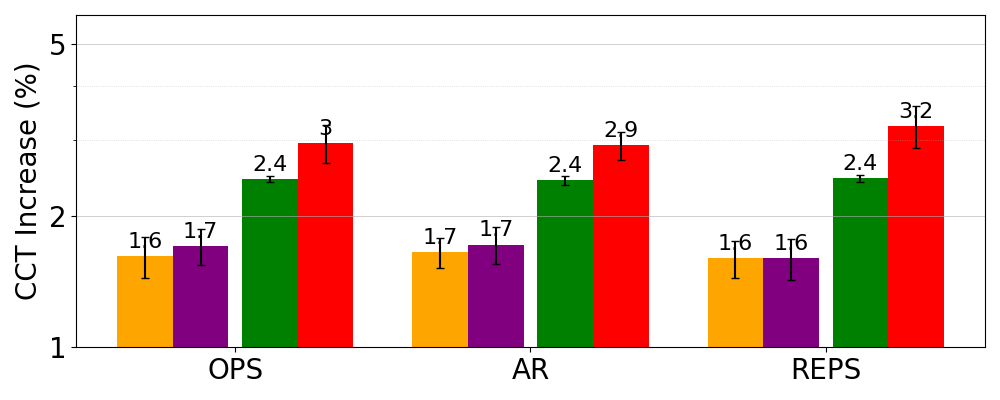}
\caption{Incast}
\label{fig:5}
\ReduceVSpace %
\end{figure}

\T{Incast.} \Cref{fig:5} illustrates the results of the incast workload. We can see that the CCT increase is relatively small, \ie the CCAs handle incast well. In addition, while the median-based algorithms do not help, they are only slightly worse than the original algorithms. Using a median with outdated information does not penalize too much.

\section{Related Work} 
A major obstacle to per-packet load balancing 
is that different paths may experience varying latencies, resulting in packet reordering that triggers congestion avoidance mechanisms and reduces throughput. This phenomenon has been extensively studied for TCP, with numerous solutions proposed over the years. Leung \etal~\cite{reordering2007overview} provide an overview of those TCP reordering solutions and discuss them in detail. A first approach~\cite{tcp-door} either temporarily disables congestion control or performs recovery during congestion avoidance upon reordering. A common technique~\cite{Improving_TCP_performance_in_multipath_packet_forwarding_networks, Enhancing_TCP_performance_to_persistent_packet_reordering} is to readjust the duplicate ACK threshold. Another  mechanism~\cite{tcp-dcr} is to delay the congestion response for one RTT after receiving the first duplicate ACK. It is also possible to readjust RTO timer calculations~\cite{tcp-rr}. More recently, it has been suggested to use network coding~\cite{network_coding_under_packet_reordering}. Another approach is to  completely ignore reordering as a loss signal and rely solely on an RTO timer with adjusted parameters~\cite{fastRTO}. Finally, Lazy TCP~\cite{LTCP} was specifically designed for datacenters. It waits for reordered packets and decreases the window only upon consecutive packet reordering. 
Unfortunately, %
these solutions do not provide solutions for %
congestion signals beyond reordering.

Congestion signals in the literature include ECN, delay, and INT signals. %
For example, DCTCP~\cite{dctcp} uses both loss signals and ECN marking. Swift~\cite{swift} employs loss signals and incorporates the delay of the last received packet. \smartt~\cite{fastflow} uses only packet trimming and timeouts as loss signals, ignoring reordering altogether, additionally relying on delay and ECN feedback. Poseidon~\cite{poseidon} mainly relies on INT signals representing the max per-hop delay (MPD) and on loss signals. DCQCN~\cite{dcqcn} relies on ECN marking, relayed back through congestion notification packets (CNPs), together with hop-by-hop, link-layer Priority-based Flow Control (PFC).

Load balancing solutions such as Hermes~\cite{hermes} try to partially avoid this mismatch of CCA and
per-packet load-balancing interaction by limiting the per-packet load balancing rerouting policy. However, they typically yield worse results than packet spraying.

Finally, CCA approaches such as MLTCP~\cite{rajasekaran2024mltcp} are orthogonal to the focus of this paper, as they look at how to converge to a round-robin schedule of concurrent job collectives.

\section{Conclusion}
As packet spraying becomes widely deployed in datacenter networks, its interaction with CCAs becomes critical, since it causes a throughput collapse for many modern CCAs.
In this paper, we modeled the throughput collapse for many CCAs when some of the paths are congested. 
Our model emphasized the crucial importance of not relying only on the latest congestion signal in these CCAs. 
Instead, we proposed using the median feedback as a more robust alternative. Building on this insight, we introduced \name and \mn, and showed how they can improve the use of packet spraying by  their baseline CCAs, while dealing with congestion in the network.

\section*{Acknowledgment} This work was partly supported by the Louis and Miriam Benjamin Chair in Computer-Communication Networks. 

\ifacm %
  
    \ifacmart 

      \bibliographystyle{ACM-Reference-Format}
      \bibliography{mybib}
    \else
      \bibliographystyle{abbrv}
      \bibliography{mybib}    
    \fi

\else
	\ifusenix %
    	{\footnotesize 
        \bibliographystyle{acm}
        \bibliography{mybib}
        }
    \else
        \ifhotnets %
            \bibliographystyle{abbrv} 
            \begin{small}
                \bibliography{mybib}
            \end{small}
          \else %
            \bibliographystyle{IEEEtran}
            \bibliography{mybib}%
        \fi %
    \fi %
\fi %

\appendices

\section{Throughput collapse theorems} \label{sec:proofs}

\T{General assumptions.} 
We follow the Mathis model assumptions~\cite{mathis} 
that 
the flow avoids retransmission timeouts, always has sufficient receiver window and data to send, the receiver is acknowledging every packet without delay, multiple congestion signals within one round trip are treated as a single one,  
and %
we can neglect rounding of the congestion window (cwnd) 
upon reduction, cwnd limits, and details of data recovery and retransmission. 
We supplement these with additional assumptions to be able to derive a closed-form model. We assume that the buffer size is large enough that the loss probability is negligible, that the congestion window is sufficiently large  that (1)~after every packet traversing the long path, three packets from short paths return first, \ie congestion causes three duplicate ACKs, 
and (2)~the congestion window does not stall while waiting for ACKs of long-path packets. 
We also assume that $n/W$ is large enough that multiple window increases occur between consecutive %
decreases. 
We further assume for all models that all packet sizes are the maximum segment size (MSS).

\old{Due to space reasons, we outline the proof for each CCA, %
and do not provide a detailed background on its mechanisms.}

\T{TCP model.} For TCP New Reno, we only make the above general assumptions.
\bp[Proof of \cref{tcp theorem}]
Each packet traversing the %
congested path returns an ACK only after ACKs from its subsequent three packets arrive, creating three duplicate ACKs and triggering a congestion window reduction by half. This observation demonstrates that packets traversing longer paths produce the \textit{exact} same window size effect as dropped packets in the Mathis model, generating an identical sawtooth pattern. Thus, our TCP flow \textit{exactly} mimics a Mathis TCP flow of RTT $\Tshort$ that periodically experiences a loss every $n$ packets. The throughput 
follows the Mathis formula.
\ep

\T{DCTCP model.}
For this model of DCTCP~\cite{dctcp}, we assume that congestion is localized at one buffer, and distinguish the result based on whether the added long-path congestion delay ($\Tlong-\Tshort$) is below the ECN buffer threshold (our default assumption) or not\new{, If it's above, we further assume that the delays $\Tlong$ and $\Tshort$ are such that ECN and reordering signals reach the sender with at least one RTT of separation.}. Note that the ECN threshold is often set around 1\,BDP~\cite{zhu2015tuning}. 
Both cases yield an $O\para{1/\sqrt{q}}$ throughput, with the same main conclusions on throughput collapse.

\bp[Proof of \cref{dctcp theorem}]
DCTCP uses both packet loss and ECN %
as congestion signals. If \textit{(i)}~ECN marking is not activated, its behavior is identical to TCP. Else, \textit{(ii)}~in addition to the reordering-driven window reductions we will have ECN-driven window reductions, yielding even worse performance. \ep

\T{Swift model.} 
For this model of Swift~\cite{swift}, we assume that there is no endpoint congestion, that $\Tshort<\targdel$\new{, $\Tlong<2\Tshort$} and that we don't reach pacing (\ie $cwnd \geq 1$).
\bp[Proof of \cref{original swift theorem}]
The Swift loss recovery is based on two main mechanisms: selective ACK (SACK) for fast recovery, and a retransmission timeout (RTO). SACK is implemented using a  sequence number bitmap. When a packet is detected as lost via a hole in the bitmap, it is retransmitted, and the congestion window is reduced multiplicatively by $1-\mm$ (where $\mm$ stands for $ max\_mdf$). 
Thus, every packet traversing a long path will cause reordering, creating  a hole in the SACK bitmap, causing a cwnd reduction. 
It effectively creates a periodic sawtooth pattern with an additive increase (AI) parameter $ai$ per entire cwnd, multiplicative decrease (MD) parameter $1-\mm$, and a cycle of $n=1/\problong$ packets. 
Let the maximum value of the cwnd be $W$ packets. The minimal value after a decrease is $W(1-\mm)$. The time for an entire cycle will be $\Delta =\frac{W\cdot \mm}{ai}$ RTTs, \ie $\Delta=\frac{W\mm}{ai}\Tshort$\, sec. 
Thus the total data delivered is the area under the sawtooth,
$$  \Delta \cdot 1/2 \cdot (W+ W\cdot(1-\mm) ) = \frac{W^2\cdot \mm \cdot(1-\frac{1}{2}\mm)}{ai}.$$
This total cycle data also equals $\frac{1}{\problong}$ by definition.
Therefore we get
$ W = \sqrt\frac{ai}{\problong\cdot(1-\frac{1}{2}\mm)\cdot\mm} $.
Since the throughput is the ratio of the cycle data by $\Delta$, combining with the formula of $\Delta$ yields the result.
\ep

We now present two models of CCAs that are reordering-insensitive, and show that they follow \textit{the same} $\Theta(1/\sqrt{q})$ throughput-collapse rule as reordering-sensitive CCAs. This emphasizes that \textit{reordering-only solutions are insufficient}.

\T{\smartt model.}
\smartt~\cite{fastflow} uses several layers of congestion control, so it is hard to model. 
Moreover, its CCA is strongly coupled to an adaptive load-balancer usage, which makes it challenging to disentangle both. 
Specifically, \smartt's ``wait to decrease" mechanism prevents congestion window reductions unless the moving average of ECN-marked packets per RTT exceeds 25\%, and relies instead on an adaptive load-balancer to handle lighter congestion. Since we assume 
round robin packet spraying, 
we always apply window reductions. %
We assume that  packets traversing the congested path experience high delay $\Tlong \geq 2 \cdot targetRTT$ and get ECN-marked, that packets traversing the short paths do not get ECN marked, and as mentioned previously, that packet drops can be neglected.
We follow suggested parameters: $targetRTT = 1.5 \cdot baseRTT$, $mi=2$, $md=2$, $fi=0.25$, $fd=0.8$, and define $\alpha$ such that $\Tshort=\alpha \cdot baseRTT$. We assume $1<\alpha<1.5$, and that $\alpha$ is big enough so that the Fast Increase component does not activate. 
We note that \smartt's multiplicative increase and decrease appear to be effectively additive, while fair decrease appears to be multiplicative.

\bp[Proof of \cref{FastFlow theorem}]
\smartt's multiplicative and fair increase mechanisms produce an additive increase of $\frac{3-2\alpha}{\alpha}$\,MSS and then another additive increase by $\frac{1}{4}$\,MSS for a total of $A \equiv \frac{12-7\alpha}{4\alpha}$\,MSS per congestion window worth of packets traversing the short paths.  
The multiplicative and fair decrease mechanisms produce a 1\,MSS decrease and then another multiplicative decrease by $\Gamma \equiv 1-\frac{fd}{BDP}$ per congested-path packet, where $BDP$ is a \smartt constant. %
Let the maximum value of the cwnd be $W$ packets. The minimal value after a decrease will be $(W-1) \cdot \Gamma$. The time for an entire cycle will be $\Delta=\frac{W-(W-1)\Gamma}{A}$ RTTs, which are $\Delta=\frac{W-(W-1)\Gamma}{A} \cdot T_{avg}$ seconds where $T_{avg}$ is the average RTT. 
The total data delivered is the area under the sawtooth $$  \Delta \cdot 1/2 \cdot (W+ (W-1)\cdot \Gamma ) = \frac{W^2\cdot (1-\Gamma^2) + 2\Gamma^2 \cdot W -\Gamma^2)}{2A}.$$ This total cycle data also equals $1/\problong$ by definition. Thus %
$W=\frac{-\Gamma^2+\sqrt{\Gamma^2 \cdot (1-\frac{2A}{\problong})+\frac{2A}{\problong}}}{1-\Gamma^2}$ and after dividing the cycle data by $\Delta$, %
{\footnotesize
\begin{equation*}
    throughput = 
    \frac{A \cdot MSS}{\problong \cdot (\Gamma+(1-\Gamma)\cdot \frac{-\Gamma^2+\sqrt{\Gamma^2 \cdot (1-\frac{2A}{\problong})+\frac{2A}{\problong}}}{1-\Gamma^2}) \cdot T_{avg}} 
\end{equation*}
}
Finally, when $\problong \to 0$, $throughput = \Theta(1/\sqrt{\problong}).$
\ep

\T{Reordering-resilient Swift  model.}
We develop a model for a \new{perfect }reordering-resilient Swift.
\new{For this model, all assumptions from the regular Swift model are maintained, with the exception of $\Tlong<2\Tshort$. We assume that $\Tlong>\targdel$ and that $\frac{fs\_\alpha}{\sqrt{W}}$ is small compared to $\targdel$ where $W$ is the maximal cwnd before a reduction.}

\bp[Proof outline for \cref{reordering-resilient Swift theorem}]
The model identifies two key interdependencies between $W$ and $\targdel$: Swift's delay-driven multiplicative window reduction depends on $\targdel$, making the steady-state maximum window $W$ a nonlinear function of $\targdel$. Conversely, Swift's flow-based scaling of $\targdel$ creates a dependency where $\targdel$ becomes a non-linear function of $W$.
To solve this coupled equation system, we employ \textit{perturbation theory} to derive an approximate model. The perturbation approach treats the coupling between $W$ and \targdel as a small deviation from the base system, allowing us to systematically improve our approximation through successive orders. \old{For space reasons, we leave out the  %
derivation details.} %
\ep

\T{Median-based CCA model.} 
\bp[Proof of \cref{thm:median}]
The congested path shifts the ACK answer for all of its packets by $\Tlong-\Tshort$ when compared to their expected time. Thus, between two congested ACKs, there are always $n-1$ uncongested ACKs with $n=1/q>3$ (because after dividing by $\Tshort$, we reduce the problem to the following: an open interval of length $n$ with non-integer endpoints always contains exactly $n$ consecutive integers, and therefore exactly $n-1$ integers that are not multiples of $n$). In the worst case, in the median window, we start and finish with congested ACKs, \eg 1 congested ACK, $n-1 \geq 3$ uncongested ones and 1 congested ACK again. Thus, we always have a majority of uncongested ACKs.
\ep
\end{document}